\documentclass[prl,preprint,preprintnumbers,amsmath,amssymb,showpacs]{revtex4-1}

\usepackage{graphicx}% Include figure files
\usepackage{dcolumn}% Align table columns on decimal point
\usepackage{bm}% bold math
\usepackage{color}
\begin{document}

%\preprint{APS}

\def\Ef{$E_{\rm F}$}
\def\Ed{$E_{\rm D}$}
\def\Eg{$E_{\rm g}$}
\def\Efmath{E_{\rm F}}
\def\Edmath{E_{\rm D}}
\def\Egmath{E_{\rm g}}
\def\Tc{$T_{\rm C}$}
\def\kpara{{\bf k}$_\parallel$}
\def\kparamath{{\bf k}_\parallel}
\def\minuskpara{$-{\bf k}_\parallel$}
\def\kparazero{{\bf k}$_{\parallel,0}$}
\def\minuskparazero{$-{\bf k}_\parallel,0$}
\def\kperp{{\bf k}$_\perp$}
\def\invA{\AA$^{-1}$}
\def\Kbar{$\overline{\rm K}$}
\def\Gbar{$\overline{\Gamma}$}
\def\Mbar{$\overline{\rm M}$}
\def\GbarMbar{$\overline{\Gamma}$-$\overline{\rm M}$}
\def\GbarKbar{$\overline{\Gamma}$-$\overline{\rm K}$}
\def\DeltaEf{$\Delta E_{\rm F}$}
\def\DeltaEfmath{\Delta E_{\rm F}}
\def\BiTe{Bi$_2$Te$_3$}
\def\BiSe{Bi$_2$Se$_3$}
\def\BiSnTe{(Bi0.67\%Sn)$_2$Te$_3$}
\def\SbTe{Sb$_2$Te$_3$}
\def\MS{\textcolor{red}{}}

\title{Reversal of the circular dichroism in the angle-resolved photoemission from \BiTe}
% sign change
 
  \author{ M. R. Scholz,$^{1,*}$ J. S\'anchez-Barriga,$^1$  J. Braun,$^2$ D. Marchenko,$^1$ 
A. Varykhalov,$^1$  M. Lindroos,$^3$ Yung Jui Wang,$^4$ Hsin Lin,$^4$ A. Bansil,$^4$
J. Min\'ar,$^2$ H. Ebert,$^2$ A. Volykhov,$^5$ L. V. Yashina,$^5$ and O. Rader$^1$ }

\affiliation{$^1$Helmholtz-Zentrum Berlin f\"ur Materialien und Energie, 
Elektronenspeicherring BESSY II, Albert-Einstein-Str. 15,  12489 Berlin, Germany}

\affiliation{$^2$Department Chemie, Ludwig-Maximilians-Universit\"at M\"unchen, 
Butenandtstr. 5-13,  81377 M\"unchen, Germany}

\affiliation{$^3$Institute of Physics, Tampere University of Technology, P.O. Box 692, 33101 Tampere, Finland}

\affiliation{$^4$Physics Department, Northeastern University, Boston, Massachusetts 02115, USA}

\affiliation{$^5$Department of Chemistry, Moscow State University, 
          Leninskie Gory, 1/3, 119992 Moscow, Russia}

\begin{abstract}

The helical Dirac fermions at the surface of topological insulators show a strong circular dichroism 
which has been explained  as being due to either the
initial-state spin angular momentum, the initial-state orbital angular momentum, or the handedness of the experimental
setup. All of these interpretations conflict with our data from \BiTe\ which depend on the photon
energy and show several sign changes. Our one-step photoemission calculations coupled to {\it ab initio} theory
confirm the sign change and assign the dichroism to a final-state effect. 
The spin polarization of the photoelectrons, instead, remains a reliable probe for the spin in the initial state. 
%650 characters
\end{abstract}

 
%73.20.At (electr str surface states),  
%79.60.Bm Photoemission: Clean metal, semiconductor, and insulator surfaces 
%75.70.Tj Magnetic properties surfaces:  Spin-orbit effects; not: 71.70.Ej (this is bulk SOC!)
%71.15.Mb (DFT cond mat)

\maketitle

Topological insulators are characterized by an insulating bulk and a metallic surface. The gap in
the  bulk bands is due to a band inversion caused by strong spin-orbit interaction
\cite{KanePRL2005,BernevigPRL06,FuPRB2007,FuPRL2007,MoorePRB07,Roy06,Murakami}. Typical examples for
which these gaps are rather large are the systems \BiSe\ and \BiTe. The metallic surface states are
protected by time-reversal symmetry \cite{FuPRL2007,MoorePRB07,Roy06,Murakami} and display
the linear $E(k)$ dispersion of Dirac fermions as demonstrated by angle-resolved photoemission spectroscopy
(ARPES) \cite{XiaNatPhys09,ChenScience09}. 
By the same method, the surface states prove robust \cite{Plucinski,King,Benia,Hirahara11,Valla,ScholzPRL12}.
The peculiar spin topology is of helical type 
and can also be resolved by ARPES when spin polarimetry is added to the experiment \cite{DilReview}.

However, spin-resolved ARPES is a rather demanding detection method since the spin polarimetry reduces the ARPES
signal by 2 to 3 orders of magnitude. On the other hand, there are many interesting subjects that require
the measurement of the spin polarization and momentum-dependent spin direction. For example, the questions
whether the in-plane spin component remains locked perpendicularly to the momentum in a warped Dirac cone
or is in-plane or out-of-plane tilted  \cite{LiangFuPRL09,DilReview,Souma11}, what the maximum value
of the polarization is \cite{Louie,Hsieh09,ScholzSARPES,Pan11,Jozwiak11}, and how the spin behaves in the
subsurface region \cite{Henk12} and during hybridization with bulk states \cite{ErmeevNatCom12}.

It is, therefore, attractive to search for an alternative method of investigation that gives the same information. 
In photoemission of core levels, the spin resolution is largely complementary to circular dichroism
in which the photoemission intensities for excitation with left- and right-circularly polarized radiation,
$I_L(E)$ and $I_R(E)$, respectively, are evaluated as a function of binding energy $E$. The data from
both methods  can be
analyzed straightforwardly with the same atomic model, as the example of spin-split $5p$ core levels
of Gd and Tb 
shows \cite{Carbone,Starke}. Also in the valence band, a circular dichroism effect in ARPES is present and
can be related to the electron spin in ferromagnetic transition metals \cite{Bansmann,Kuch}.

We have recently observed a strong circular dichroism effect in ARPES from the topological surface state of
\BiTe\ \cite{ScholzSARPES}. Spin-resolved ARPES from the same system excited by linearly polarized light has
been compared to the circular dichroism in ARPES without spin resolution. This showed that the circular dichroism
reverses together with the spin texture at the binding energy of the Dirac point \cite{ScholzSARPES}. 
The circular dichroism asymmetry $A=(I_L-I_R)/(I_L+I_R)$ was found to be very large ($>20$\%\ \cite{ScholzSARPES})
when  compared to measurements from ferromagnets ($\sim3$ to $5$\%\ \cite{Bansmann,Kuch}). 

For \BiSe, a strong spin polarization of the topological surface state has been observed by spin-resolved ARPES 
\cite{Pan11,Souma11,Jozwiak11}, and in addition several measurements of the circular dichroism in ARPES have
been reported for this system \cite{Wang11,SRPark11,IshidaPRL11}. Wang {\it et al.} conducted circular dichroism
measurements at a photon energy of $h\nu=6$ eV using a pulsed laser source and a time-of-flight detector \cite{Wang11}.
The dichroism effect is interpreted assuming transitions into a spin-degenerate continuum of final states and,
therefore, is sensitive only to the spin in the initial state of the photoemission process. By measuring one spin
component in the surface plane under two different angles 
and applying symmetry arguments, the two in-plane spin components $\left< S_x \right>$,
$\left< S_y \right>$ and the perpendicular component $\left< S_z \right>$ are determined \cite{Wang11}.
Another measurement on \BiSe\ at photon energies of 10 and 13 eV   led to similar results and a circular dichroism
effect of 30\%\ \cite{SRPark11}. Park {\it et al.} concluded that for these photon energies a free-electron final
state can be assumed and for left- and right-circularly polarized light final states of different orbital angular
momentum character are reached. The orbital angular momentum was found to be locked to the momentum in a similar
way as the spin, and the orbital and spin angular momenta were determined to be antiparallel to each other
\cite{SRPark11}. Jung {\it et al.} reached the same conclusion of antiparallel orbital and spin 
angular momenta for \BiTe\ \cite{Jung11}. Ishida {\it et al.}
compared Cu-doped \BiSe\ to SrTiO$_3$ for which a dichroism effect $>60$\%\ is observed \cite{IshidaPRL11}.
The measured circular dichroism of Cu$_x$\BiSe\ at $h\nu\sim 7$ eV appears similar to that reported in the other
studies but is assigned to a geometrical origin. In the corresponding experimental setup there are two planes
perpendicular to the sample surface for which the dichroism disappears \cite{IshidaPRL11}. This is first of all
the detection plane spanned by the incident light and the photoelectron momentum. Furthermore, another plane
normal to the surface and perpendicular to the detection plane is identified where the dichroism disappears as well
because of the wave function symmetry. This plane is a nodal plane only for circular dichroism of the topological
surface state due to a mirror symmetry of its effective Hamiltonian and its two-dimensionality, in contrast to the
bulk valence band states. Accordingly, it can be seen as a specific feature reflecting the two-dimensionality of
the electronic states \cite{IshidaPRL11}.

In the present work, we investigate the circular dichroism in ARPES of \BiTe\ for different photon energies and
demonstrate that our data question the previous interpretations. It is concluded that current explanations are
too simple and a more detailed theoretical treatment is required. Corresponding calculations are conducted and 
compared to the experiment.

We have grown single crystals of \BiTe\ by the Bridgman method and cleaved them {\it in situ}. The achieved (111)
surfaces are of high quality as concluded from the sharp features in angle-resolved photoemission of the valence
band (Fig. 1). Measurements have been carried out in ultrahigh vacuum of $1\times10^{-10}$ mbar 
at low temperature (30--40 K) with a Scienta
R8000 electron analyzer at the UE112-PGM2a beam line of BESSY II with circularly-polarized undulator radiation.
The geometry of the experiment is shown in Fig. 1.

The bulk electronic structure of \BiTe\ is obtained by performing first-principles calculations within the 
framework of the density functional theory using the generalized gradient approximation to model exchange-correlation
effects \cite{LindroosPRB02}.
The spin-orbit coupling is included in the self-consistent cycles of the electronic structure calculation. 

The angle-resolved photoemission intensity calculations are based on the one-step model \cite{hop80}. We use a fully
relativistic formalism, allowing to consider in a natural way effects in the photocurrent calculation
induced by spin-orbit coupling because the practical calculation is based on the Dirac formalism \cite{bra96,ebe11}.
The spin-orbit coupling enters thus in the calculation of the ground 
state and again in the photoemission calculation.
The photoemission calculation itself is based on multiple-scattering theory, using explicitly the low-energy electron
diffraction (LEED) method to calculate the initial and final states for a semi-infinite atomic half-space. In this way
the final state is calculated by the best available single-particle approach as a so-called time-reversed LEED
state \cite{gray11}. In line with this, the initial state is represented by the retarded one-electron Green function
for the same semi-infinite half-space. The photoemission calculations include matrix-element effects, multiple scattering
effects in the initial and final states, the effect of the photon momentum vector, and the escape depth of the photoelectrons
via an imaginary part in the inner potential. These lifetime effects in the final states have been included in our
analysis in a phenomenological way using a parameterized complex inner potential $V_{\rm o}(E) = V_{\rm or}(E) + iV_{\rm f}(E)$. 
Herein, the real part serves as a reference energy inside the crystal with respect to the vacuum level.  To account for
impurity scattering, a small constant imaginary value of $V_{\rm i}$ = 0.004 eV was used for the initial state. 
A realistic description of the surface potential is given through a Rundgren-Malmstr\"om barrier \cite{malm80}
which connects the asymptotic regime $z<z_{A}$ to the bulk muffin-tin zero V$_{\rm or}$ by a third order polynomial in $z$,
spanning the range $z_{A}<z<z_{E}$. 
In other words, $z_A$ defines the point where the polynomial region starts whereas $z_{E}$
defines the point where the surface region ends and the bulk region starts with the first atomic layer. The effective
$z$-dependent surface barrier $V$ is scaled with respect to the vacuum level $E_{vac}=0.0$ eV utilizing the value of the
work function $\phi=5.0$ eV. The zero of the $z$ scale lies in the uppermost layer of atoms.

The geometrical origin for a circular dichroism in ARPES as has been described in detail in Ref.~\cite{Kuch} can be understood
on the basis of Fig.~1. No dichroism is expected when the geometry is such that a symmetry operation which transforms right-
into left-circularly polarized light leaves the momentum vector of the photoelectron {\bf k} unaffected. This is shown  at the
top of Fig. 1. The green plane ($\eta=0^\circ$, $k_y=0$) is the plane of incidence  and is a mirror plane transforming the 
right circularly polarized light of Fig. 1(a) into the left circularly polarized light of Fig. 1(b) without affecting {\bf k}.
The green plane constitutes, therefore, a nodal plane meaning zero dichroism for $k_x=0$. For all other
situations a dichroism can occur, for example in the yellow plane ($k_y=0$) which is the detector plane. (The detector slit
is indicated as a narrow vertical rectangle in Fig.~1.) 
Reversal of the light polarization changes {\bf k} from $+\eta$ to $-\eta$ [Fig. 1(c)] which means opposite sign of the
dichroism for emission angles $+\eta$ and $-\eta$ [Fig. 1(d)]. 
Figure 1(e)--(g) show separately the photoemission intensity for right- and
left-circularly polarized light of \BiTe. 
The effect is apparently very large and visible already comparing the intensities I$_L$ and
I$_R$. The corresponding asymmetry $A$ is then shown in a color
representation in Figs.~2--4. Figures 2--4 display a white line in their center, $k_y=0$, that represent the nodal
plane mentioned above. As can be seen, not only the topological surface state shows the nodal plane but also the bulk
states, a fact which would support the interpretation as a geometrical effect. However, it should be stressed that the additional
perpendicular nodal plane discussed in Ref. \cite{IshidaPRL11} is identical to the plane in which the data of Fig.~1(e)--(g)
and Fig. 2 have been measured. This means that the description suggested by Ishida {\it et al.} as dichroism due to geometrical
effects only is not appropriate for our present case. Moreover, the reversal of the dichroism effect at the binding energy
of the Dirac point contradicts the model by Ishida {\it et al.} because this behavior cannot be explained based on purely 
geometrical effects.
  
Figure 2 shows the behavior of the circular dichroism in the experiment at photon energies between 21 and 100 eV. 
The asymmetry is very large (80\%\ at $h\nu=55$ eV). 
(Note that, as a two-dimensional state, the topological surface state has a binding energy independent of the photon energy 
which may still change slightly  within Fig. 2 due to surface doping by residual gas.)
The top row [Figs. 2(a--d)] shows complete $E(\kparamath)$ dispersion relations which are all characterized by a 
circular dichroism effect that changes sign with binding energy at the Dirac point. Generally, the lower-binding-energy
range shows a large  dichroism effect. ($\kparamath$ is the projection of the electron wave vector on the surface
plane.)
This range contributes spectral intensity of the topological surface state only while the
higher-binding energy range has to some extent bulk-like contributions as well. We see that for each selected
photon energy in Figs. 2(a--d)  the complete circular dichroism signal has reversed sign, above as well as below
the Dirac point. Figure 2(e) shows 
for $h\nu<70$ eV the
detailed behavior with smaller steps in photon energy. 
The behavior around 45 eV may indicate additional sign changes \cite{supplement}. 
According to the initial-state model for the spin \cite{Wang11},
the data in Fig. 2 would mean that 
the spin of the probed ground state changes during the photon energy scan. This is not possible,
not even  under the assumption of a layered spin texture, which has been predicted to reverse in the topmost atomic
layers \cite{Henk12}, because the photocurrent from the topmost layer dominates in the photoemission signal from
the topological surface state. A similar problem arises with the initial-state orbital angular momentum model
\cite{SRPark11} because, again, it is unclear how the probed initial state can depend on the photon energy of the
probing radiation. 
 
We have, therefore, performed calculations for the ground state and calculated the photoemission spectra as described briefly
above. Figure 3(a) shows the resulting dichroism in ARPES at $h\nu=27$ eV using the same representation as for the
experimental data. As can be seen, the angle- and the binding-energy-dependences are very similar to the experiment
in Fig. 2. For investigating the origin of the dichroism effect further, our photoemission calculations allow us  to vary the 
spin-orbit-coupling strength. The idea behind this is that the decoupling of spin and orbital moment will reduce
the effect that the spin polarization can have on the circular dichroism. A reduced spin-orbit
coupling can be simulated simply by an increased speed of light $c_0$. 
Before applying this in Fig. 3, we tested the use of $1.09c_0$ and $1.24c_0$ 
on the spin-orbit splitting of W metal 
between $\Gamma_{7+}$ and $\Gamma_{8+}$ states and obtained   90\%\ and 75\%, 
respectively, of the original splitting. 
This effect is not too far from the expected scaling with $1/c_0^2$ in the atom.
The first impact that one notes   in Fig. 3  is the change in the binding
energy of the Dirac point.  This is due to the effect of the spin-orbit coupling on the bulk band inversion. It should be
stressed that if the spin-orbit coupling would be reduced further the band inversion and with it also the topological
surface state would disappear. Figures 3(b) and 3(c) show that the changes of the circular dichroism of the 
topological surface state are rather small when the spin-orbit coupling is reduced. 
This  suggests  that the contribution of the spin to the circular dichroism is a minor one, and we will return to this
question further below in connection with the photon-energy dependence. 

One-step-photoemission intensities in general are based on ground state electronic structure calculations
and, as a consequence, the energetics of the final states obtained for higher photon energies often deviates from the
experimental situation depending on the excitation energy. In this sense, our calculations have qualitative
character and do not, e. g., reproduce the position of the experimental sign change between 21 and 25 eV photon
energy. However, Figure 4 shows that between 25 and 50 eV the sign has clearly reversed in the calculation.
This is an important confirmation of the experimental results of Fig. 2 demonstrating that the sign change 
is reproduced by our photoemission theory and can  be ascribed to
the final states. 

We have previously
investigated the L-gap surface state of Cu(111) by 
one-step photoemission calculations for a wide photon-energy range from 21 to 70 eV and compared to ARPES
experiments \cite{Mulazzi09}. 
In that case, the dichroism asymmetry depends on the photon energy as well. An
analysis of the final states confirmed that the dichroism effect is strong where a $d$-type final state is reached
in agreement with the expectations from selection rules for the orbital angular momentum
\cite{Mulazzi09}. 
This mechanism has been discussed later on in a very similar context in an ARPES study on Cu(111) and Au(111)
\cite{BKimCuAu12}. 
The same mechanism underlies the present results which also involve transitions from $p$-type
initial states to $d$-type final states. 
When the present calculation is modified to exclude transitions into $d$-type final states, 
      the sign change between 25 and 50 eV photon energy disappears \cite{supplement}. 
The final states are, however, more difficult to analyze than in fcc Cu 
due to frequent backfolding because of the small size of the bulk Brillouin zone along $z$ (perpendicular to the surface). 
An analysis of the initial state identifies the topological surface state as being due to all three $p$-orbitals,
and this is the reason for the stronger dependence on the final states concerning the sign changes as compared to
the $p_z$-type surface state of Cu(111). The spin-orbit coupling is much weaker in Cu when compared to the present
system \BiTe. The fact that spin-orbit coupling is not a precondition for a strong circular dichroism in ARPES 
has been demonstrated also early on: In  graphite,  transitions from the $\pi$-band into $d$ final states lead to
dichroism asymmetries of up to 60\%\ at a negligibly small spin-orbit coupling \cite{SchoenhenseEPL92}.
It should be stressed that spin-orbit interaction also induces a $\bf k$-dependent spin-polarization of the initial states,  
simply due to the presence of the surface \cite{HeinzmannDil12}. This effect, which is quantitatively considered in our analysis, has to be accounted for
to fully understand the experimental data.

Figure 4 also allows us to inspect the potential of spin-resolved photoemission and its depencence
on the final states. The bottom row of Fig. 4 shows the calculated spin-polarization 
$P=(I^\uparrow-I^\downarrow)/(I^\uparrow+I^\downarrow)$ of the photoelectrons as a function
of the photon energy for linearly polarized light. 
The sign of the spin polarization remains constant and also its value is practically the same
($P=80$\%\ at 25 eV and 75\%\ at 50 eV when evaluated 100 meV above the Dirac energy). This means that spin-resolved
photoemission is much less affected by final-state effects than the circular dichroism and, therefore, can be used
to deduce the electron spin in the initial state rather directly.
This has to do with the fact that the circular dichroism is the 
difference of two signals and by itself very sensitive
while for the initial-state spin the phase space for excitation
by linearly polarized light is less restricted.

In summary, we have investigated the dependence of circular dichroism in ARPES from \BiTe\ on the photon energy. Even
for final-state energies that are 10 times higher than those considered free-electron like, we observe a sensitive
dependence on the photon energy and a reversal of the sign. The various proposed initial-state models favoring the spin
or the orbital angular momentum cannot be applied, and the same holds for pure geometric models. The circular dichroism
in ARPES has clearly been identified as a final-state effect experimentally with confirmation by the results of one-step
photoemission calculations. The spin polarization of the photoelectrons is not affected.  

O. R. thanks G. Bihlmayer for helpful discussions. 
Financial support from the Deutsche Forschungsgemeinschaft (Grant No. EB-154/18) and the 
Bundesminsisterium f¨ur Bildung und Forschung (Grants No. 05K10WMA) is gratefully acknowledged.

{\it Note added:} Recent one-step photoemission calculations \cite{Henk12b} do  not 
support the conclusions by Jung {\it et al.} \cite{Jung11}.
A normal-incidence geometry is suggested
as probe for the initial-state spin perpendicular to the surface  \cite{Henk12b}.

$^*$Present address: Physikalisches Institut, Universit\"at W\"urzburg, Am Hubland, 97074 W\"urzburg, Germany.

\newpage

\noindent FIG. 1. (Color online)  (a-d) Experimental photoemission geometry and comparison between \BiTe\ data along 
the \Gbar-\Kbar\ direction excited by (e) linearly and (f) right and (g) left circularly polarized light
of $h\nu=55$ eV. The topological surface state is identified by its linear dispersion upwards from the Dirac
point around 0.25 eV binding energy, and the dichroism is strongly visible without need for difference spectra.

\phantom{Zeile}

\noindent FIG. 2.  (Color online)  Change of the circular dichroism effect with photon energy.
(a-d) The upper row shows three reversals of the sign between photon energies of 21 and 100 eV.
(e) Plot of the dichroism asymmetry $A$ evaluated 100 meV above the Dirac point {\it vs.} photon energy. 
The dichroism asymmetry is large (80\%\ at $h\nu=55$ eV).

\phantom{Zeile}

\noindent  FIG. 3. (Color online) Calculations for a reduced spin-orbit interaction for the example of $h\nu=27$ eV. 
The spin-orbit coupling  is reduced from (a) 100\%\ ($c_0$) to (b) 90\%\ ($1.09c_0$) and (c) 75\%\ ($1.24c_0$). 
This modifies the inverted
bulk band gap and moves the topological surface state in energy. The circular dichroism remains very similar
indicating a minor role of the electron spin polarization for the circular dichroism. 
The overall change in binding energy is a side effect and not relevant here.

\phantom{Zeile}

\noindent FIG. 4.  (Color online) Results from the one-step photoemission calculation. Top: The calculated 
circular dichroism changes sign between photon energies of (a) 25 eV and (b) 50 eV. 
Bottom: For the same system but linearly polarized light. The calculated spin polarization $P$ of the photoemission from the topological surface state
is unaffected by the photon energy. It reaches $P\sim80$\%\ at 25 eV in (c) and $75$\%\ at 50 eV in (d). 

\begin{figure}[ht]
	\centering
  \includegraphics[width=1\textwidth]{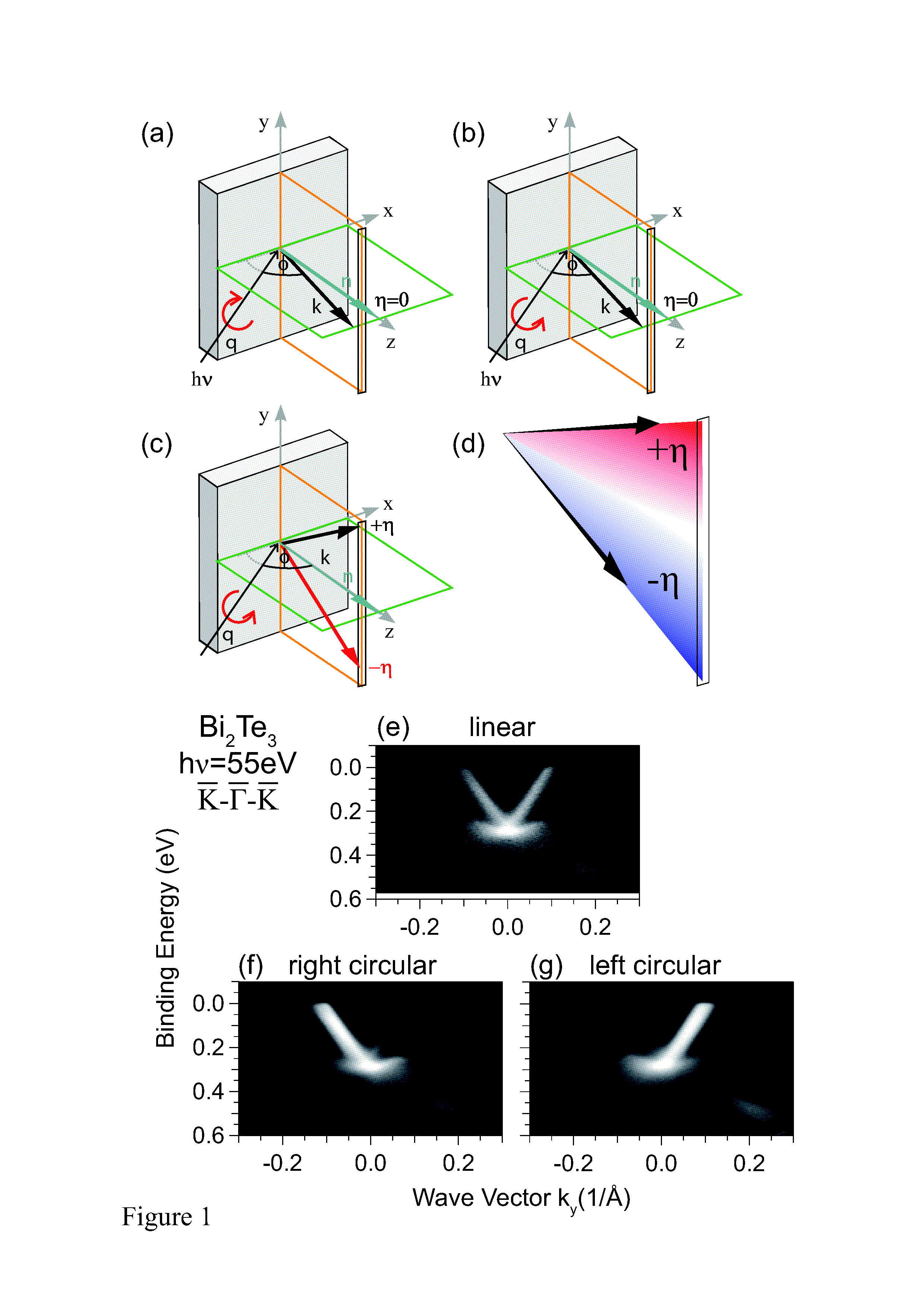}
\end{figure}

\begin{figure}[ht]
	\centering
  \includegraphics[width=1\textwidth]{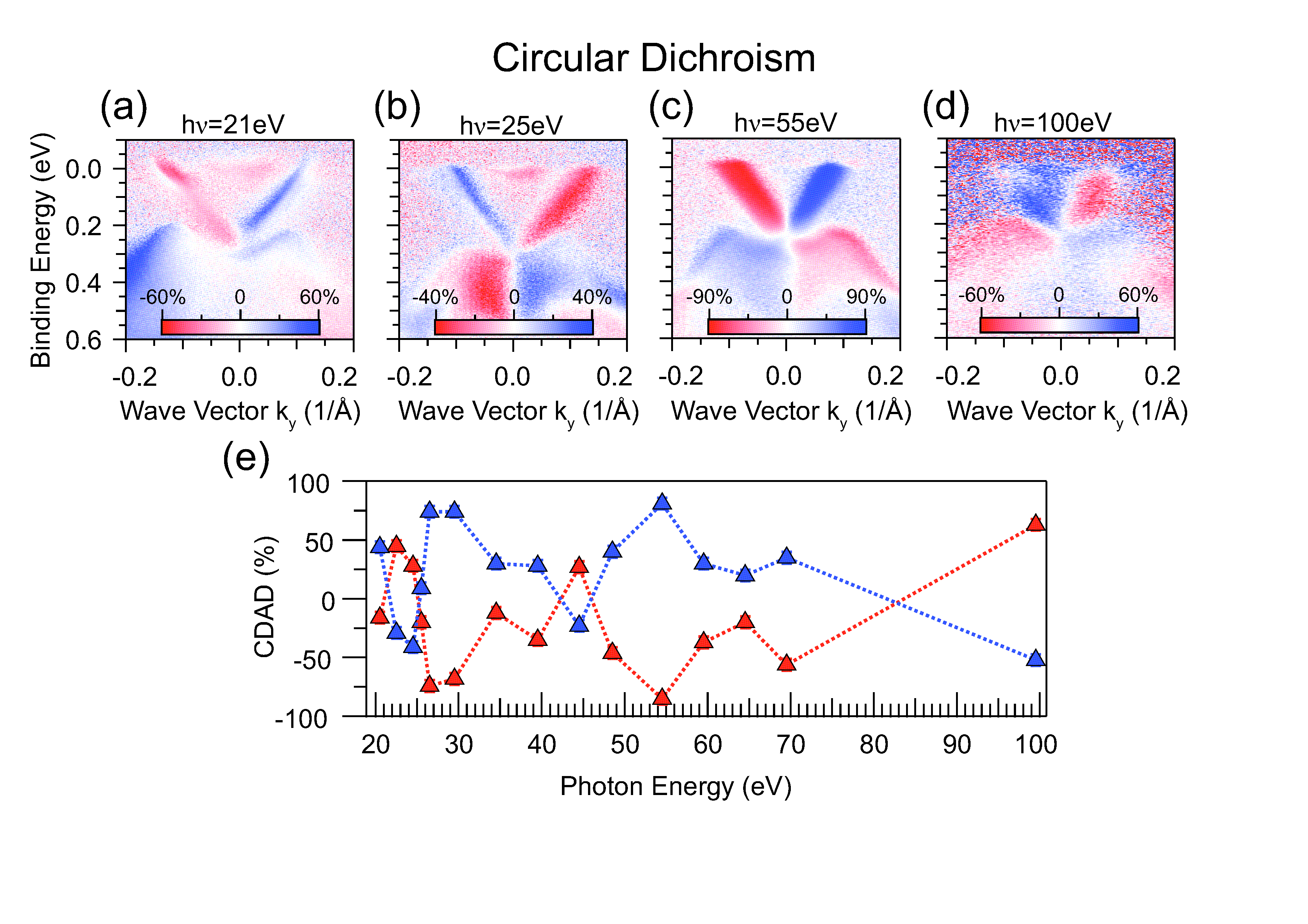}
\end{figure}

\begin{figure}[ht]
	\centering
  \includegraphics[width=1\textwidth]{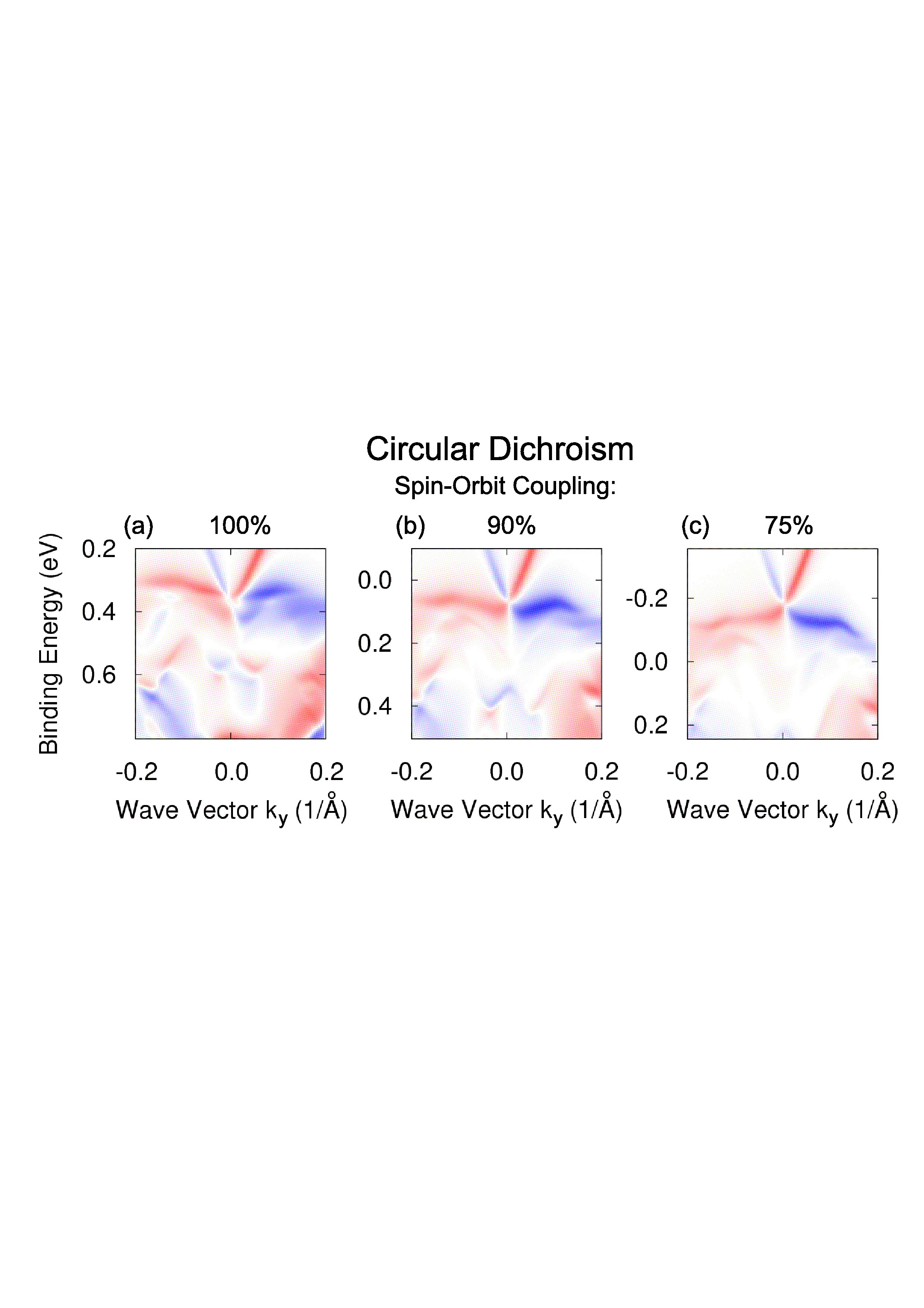}
\end{figure}

\begin{figure}[ht]
	\centering
  \includegraphics[width=1\textwidth]{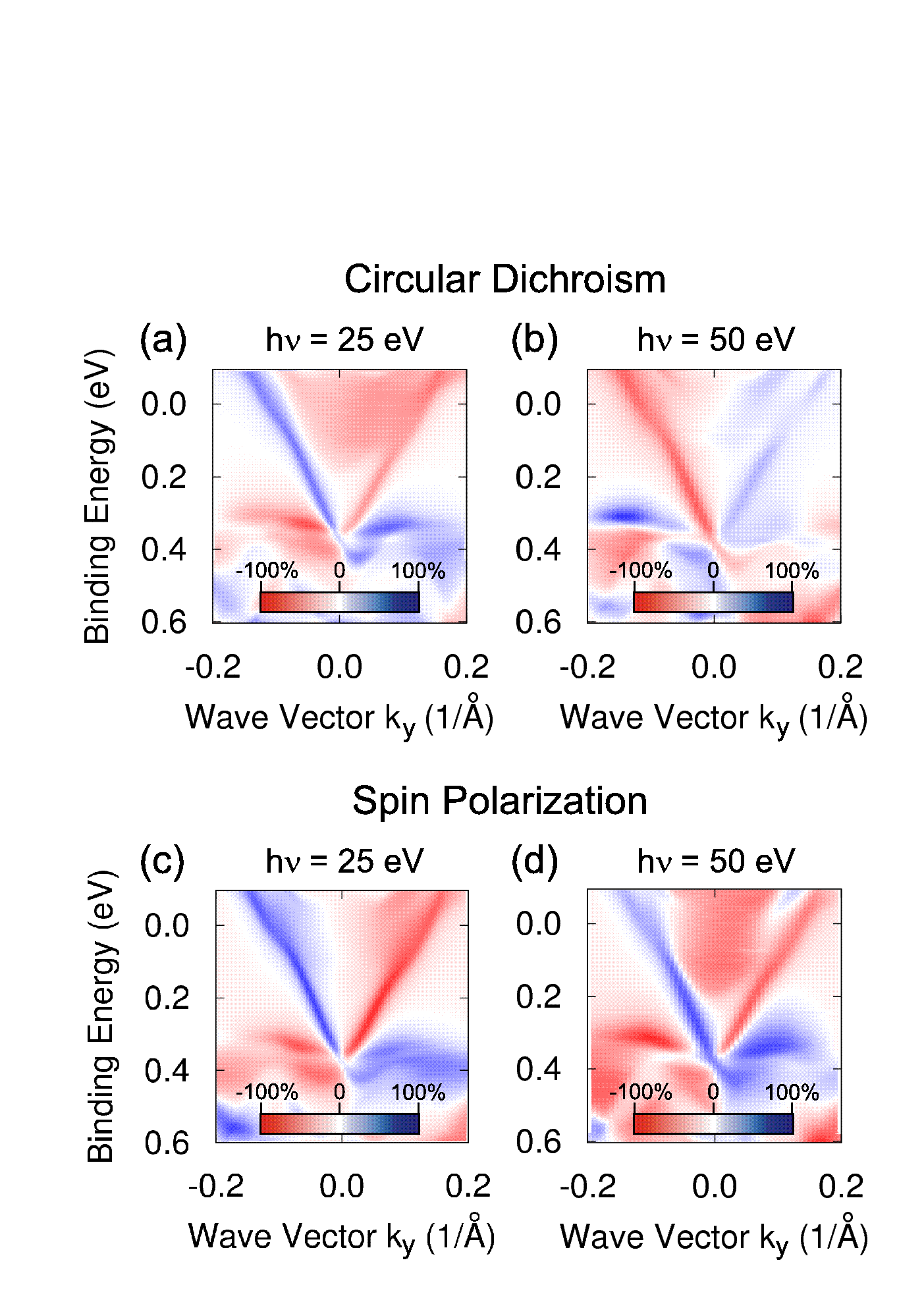}
\end{figure}
 
\end{document}